\documentclass{gfl}
\usepackage{epsfig}
\usepackage{natbib}

\newcommand{\ltaraw}{$\; \buildrel < \over \sim \;$}
\newcommand{\lta}{\lower.5ex\hbox{\ltaraw}}
\newcommand{\gtaraw}{$\; \buildrel > \over \sim \;$}
\newcommand{\gta}{\lower.5ex\hbox{\gtaraw}}

\newcommand{\rref}{}

\loadboldmathitalic \title[Conformal  Confusion]{Coordinate Confusion in
  Conformal Cosmology\footnotemark[1]} \author[Lewis et al.]{ Geraint F.
  Lewis$^{1}$, Matthew J. Francis$^1$, Luke A. Barnes$^{2,1}$
  \& J. Berian James$^{3,1}$\\
  $^{1}$Institute of Astronomy, School of Physics, A28,
  University of Sydney, NSW 2006, Australia\\
  $^2$Institute of Astronomy, Madingley Rd, Cambridge, UK\\
  $^3$Institute for Astronomy, Royal Observatory, Edinburgh EH9 3HJ, UK\\
} \date{\today}
\begin{document}
\maketitle
\label{firstpage}
\begin{abstract}
  A       straight-forward        interpretation       of       standard
  Friedmann-Lemaitre-Robertson-Walker (FLRW) cosmologies is that objects
  move  apart due  to  the  expansion of  space,  and that  sufficiently
  distant galaxies must be receding at velocities exceeding the speed of
  light.   Recently,  however,  it  has  been suggested  that  a  simple
  transformation  into  conformal  coordinates can  remove  superluminal
  recession velocities, and hence the  concept of the expansion of space
  should  be  abandoned.  This  work  demonstrates  that such  conformal
  transformations do not eliminate superluminal recession velocities for
  open   or  flat   matter-only  FRLW   cosmologies,  and   all  possess
  superluminal  expansion.  Hence,  the  attack on  the  concept of  the
  expansion  of  space based  on  this  is  poorly founded.   This  work
  concludes  by emphasizing  that the  expansion of  space  is perfectly
  valid  in  the general  relativistic  framework,  however, asking  the
  question of whether space {\it really} expands is a futile exercise.
\end{abstract}
\begin{keywords}
 cosmology: theory
\end{keywords}

\long\def\symbolfootnote[#1]#2{\begingroup%
  \def\thefootnote{\fnsymbol{footnote}}\footnotetext[#1]{#2}\endgroup} 

\def\newblock{\hskip .11em plus .33em minus .07em}
\section{Introduction}     \label{intro}     \symbolfootnote[1]{Research
  undertaken  as part  of  the Commonwealth  Cosmology Initiative  (CCI:
  www.thecci.org),  an  international  collaboration  supported  by  the
  Australian Research Council}

While   it  is  almost   a  century   since  \citet{1929PNAS...15..168H}
identified the  expansion of  the Universe, debate  is still  ongoing to
what this  expansion physically means. The mathematics  of cosmology are
set within  the framework of general relativity  and textbooks typically
describe the expansion of the  universe as an expansion of space itself.
However, while  the concept of  expanding space has recently  been under
fire \citep{2004Obs...124..174W,Peacockweb},  it is clear  what has been
attacked is  a particular  picture of space  expanding like a  fluid and
carrying galaxies along  with it; \citet{barnes2006} and \citet{Francis}
have  demonstrated that it  is correct  to talk  about the  expansion of
space,  as long  as  one  clearly understands  what  the mathematics  of
general relativity is telling us.

However, some recent attacks on the picture of expanding space have been
more  forceful  \citep[e.g.][]{2005PASA...22..287C,2006astro.ph.10590C},
with  a typical  line  of  criticism invoking  a  comparison between  an
explosion of massless particles  in static, flat spacetime (Milne model)
and     empty     FLRW     universes.      In    a     recent     paper,
\citet{2006astro.ph..1171C} examines  the nature of  FLRW cosmologies in
conformal  coordinates,  concluding   that  superluminal  separation  of
objects can be removed through a simple change of coordinates, and hence
that  superluminal expansion  is  illusionary; this  is  in contrast  to
\citet{2004PASA...21...97D},  who  point   out  that  such  superluminal
expansion is  a generic  feature general relativistic  cosmologies.  The
goal of  this short contribution  is to clear  up some of  the confusion
surrounding   the    concept   of   expanding    space   and   conformal
transformations,   showing   that   superluminal  expansion   does   not
necessarily vanish in conformal coordinates. Furthermore, the concept of
expanding space  is reasserted as  a valid description of  the universe,
although discussion on whether space  {\it really} expands is seen to be
futile.

\begin{figure}
\centerline{ \psfig{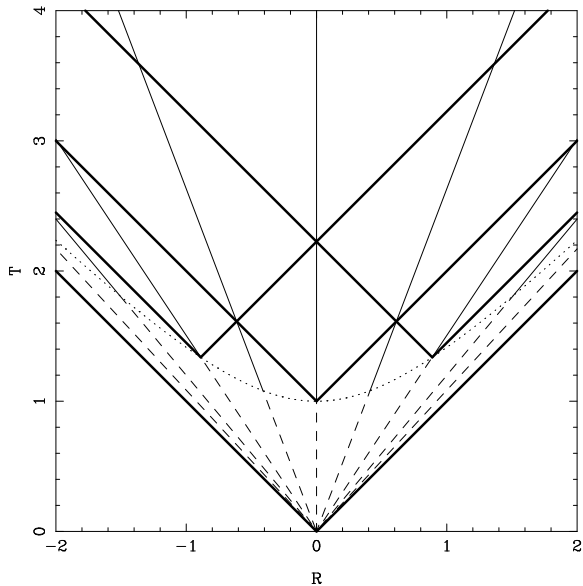}}
\caption[]{An open universe in conformal coordinates. The thick, solid
lines denote the path of light rays in conformal coordinates, whereas
the dashed and solid lines represent the paths of fundamental observers.
The entire (infinite) open universe is contained within the outermost
light cone. The dotted hyperbola represents the big-bang.
\label{fig1}}
\end{figure}

\section{General Relativistic Cosmologies}\label{gr}

\subsection{FLRW Universes}\label{universe}
The starting  point for  a standard, general  relativistic model  of the
cosmos  begins with  the assumption  of homogeneity  and  isotropy. With
this, the  spacetime of the universe  can be described by  a FLRW metric
with invariant interval of the form
\begin{equation}
ds^2 = dt^2 - a^2(t)\left[ dx^2 +R_0^2 S^2_k(x/R_0) (d\theta^2
+ \sin^2\theta\ d\phi^2) \right]
\label{flrw}
\end{equation}
where $S_k(x) =  \sin\ x, x, \sinh\ x$ for  spatial curvatures of $k=+1$
(closed),  $k=0$  (flat)  and   $k=-1$  (open)  respectively,  with  the
curvature given by $R_0^{-2}$;  note, throughout $c=1$.  Also, $a(t)$ is
the  scale factor,  whose evolution  depends  upon the  relative mix  of
energy  density in  the Universe.   It is  clear from  this form  of the
metric  given by Equation~\ref{flrw}  that for  a fixed  coordinate time
$t$, the physical separation of objects depends on the size of the scale
factor  $a(t)$, and  the  increase of  $a(t)$  with $t$  results in  the
increasing  separation of  objects; this  is typically  taken to  be the
expansion of the Universe.

\subsection{Velocities in Expanding Universes}\label{velocities}

In order  to understand  superluminal recession, we  must first  be very
clear  about how  we are  defining  recession velocity  in an  expanding
universe.  A fundamental definition of distance in general relativity is
the  proper distance,  defined  as the  spatial  separation between  two
points along  a hypersurface  of constant time.   Given the form  of the
FLRW metric  (Equation \ref{flrw}), the radial distance  from the origin
to a coordinate $x$ along a hypersurface of constant $t$ is;
\begin{equation}
D_{p}(t) = a(t)\ x
\end{equation}
Taking  the  derivative  with  respect  to  coordinate  time  [which  is
synchronous for all comoving observers  (fixed $x$) and is equivalent to
their proper time $\tau$] we obtain  what we will refer to as the proper
velocity
\begin{equation}
  v_{p}    \equiv \frac{dD_p}{d\tau}=   \frac{dD_{p}}{dt}     =    \frac{da}{dt}x    +
  a\frac{dx}{dt}
\end{equation}
For  comoving  observers with  $dx/dt=0$  this  becomes  the well  known
distance-velocity law.   However, universes which  are open or  flat are
spatially  infinite  and the  above  metric  predicts that  sufficiently
distant  objects will  separate  at velocities  exceeding  the speed  of
light; this issue has introduced  a lot of confusion and discussion into
the nature of the expansion~\citep{2004PASA...21...97D}.

The coordinate velocity can also be defined as
\begin{equation}
  v_{c} = \frac{dx}{dt}
\end{equation}
For the  FLRW metric, comoving  observers have coordinate  velocities of
zero, and  peculiar velocities $adx/dt$ must  be less than  unity, to be
consistent with  special relativity \citep[see][]{Francis}.   It follows
that  all  radial coordinate  velocities  in  the  FLRW metric  will  be
subluminal.  This reflects  a feature of the coordinate  system; what is
important however is not how arbitrarily defined coordinates change with
respect  to one  another but  how the  proper distance  between  any two
points changes with respect to the proper time of observers.

\begin{figure*}
\centerline{ \psfig{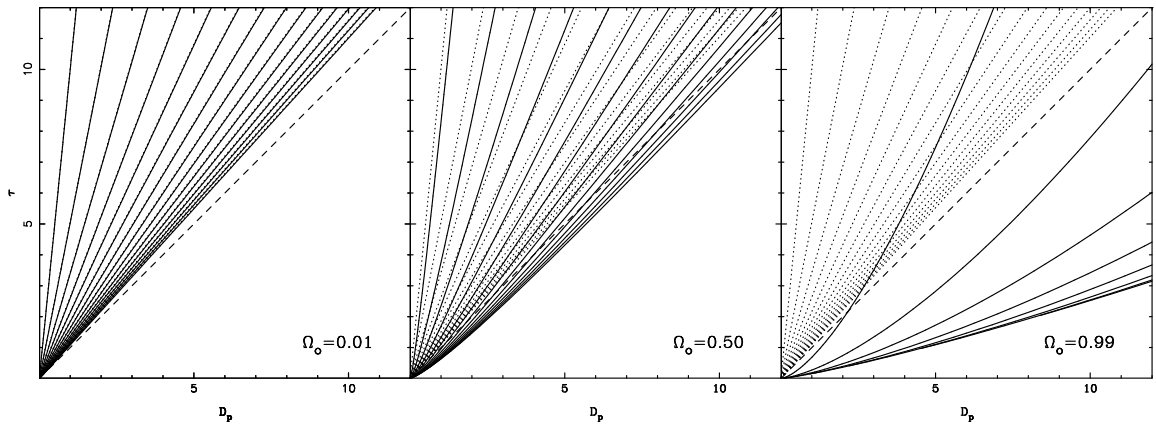}}
\caption[]{The proper distance to several comoving observers in several
  open, matter only universes (solid curves). The dashed line at $45^\circ$ 
represents the speed of light. The dotted lines represent the recession 
paths of the fundamental observers integrated over the conformal
coordinates without considering the conformal factor outside the metric. 
\label{fig2}}
\end{figure*}

\subsection{Conformal Transformations}\label{conformal}
Conformal  transformations  are important  in  understanding the  causal
structure   of  spacetime   \citep{1973lsss.book.....H}.    A  conformal
transformation  maps  from  one  set  of coordinates  to  another  while
preserving   angles  and   infinitesimal  shape,   and   two  spacetimes
represented by metrics $g'$ and $g$ are conformally equivalent just if
\begin{equation}
g'({\bf x}) = \Omega({\bf x})^2\ g({\bf x})
\label{conform}
\end{equation} 
where $\Omega({\bf  x})$ is a  scalar function~\footnote{More precisely,
  two metrics are  conformally equivalent if they possess  the same Weyl
  tensor.}   This function  can be  interpreted as  a scalar  field that
influences perfect rulers  and clocks to distort one  spacetime into the
other. A metric  that is conformally equivalent to  the Minkowski metric
is labeled `conformally flat'.

An examination of the  FLRW metric (Equation~\ref{flrw}) reveals that it
is  conformally  flat\footnote{  For  flat spacetime,  the  Weyl  tensor
  vanishes identically.   This can be simply  shown to be  true for FLRW
  spacetime using  a symbol mathematics  package such as  grtensor ({\tt
    http://grtensor.phy.queensu.ca}).}  and hence  can be written in the
form
\begin{equation}
ds^2 = \Omega^2({\bf x})\ ds^2_{\mathrm{flat}}
\label{toflat}
\end{equation}
where $ds_{flat}$  represents the  spacetime of special  relativity. The
precise  form of $\Omega({\bf  x})$ changes  depending on  whether flat,
closed or open cosmologies  are considered.  This spacetime mapping from
the FLRW  metric to the  Minkowski metric, also subsumes  null geodesics
(the  motion of  photons,  which satisfy  $ds=0$),  i.e.  the  distorted
lightcones  seen  in cosmological  coordinates  can  be  drawn onto  the
classical   lightcones  of  special   relativity  \citep[see   Figure  1
in][]{2004PASA...21...97D}.

Typically,  conformal  representations of  FLRW  universes  consider only  the
radial motion  of photons  and neglect the  angular components of  the metric.
With such  a transformation, fundamental,  or comoving, observers  (with fixed
$x$, $\theta$  and $\phi$ in  Equation~\ref{flrw}) move on  straight, vertical
lines on  an $R$-$T$ representation of  flat spacetime, while  photons move at
$45^\circ$ {\rref (the coordinate  transformation from open FLRW coordinates to
  conformal  coordinates  for  an  open  universe is  discussed  in  detail  in
  Section~\ref{open})}.  Such  an approach  has proved to  be very  powerful in
  understanding cosmic causality and the nature of fundamental horizons in the
  Universe~\citep{1956MNRAS.116..662R,1988CQGra...5..207E},   However,  it  is
  important to note that the consideration of purely radial paths results in a
  representation which is not fully conformal; the mathematical transformation
  of the  full FLRW  metric into conformally  flat coordinates was  tackled by
  \citet{is}. An important result from  their study is that in fully conformal
  coordinates, fundamental  observers (comoving observers in  FLRW metrics) no
  longer  travel along  straight, vertical  paths;  this is  examined in  more
  detail in the next section.

\subsection{An Open Universe}\label{open}
\citet{2006astro.ph..1171C}  considers  the  question of  the  conformal
representation of the  FLRW metric, focusing, as a  specific example, on
an open  universe. Starting with the  FLRW metric (Equation~\ref{flrw}),
he shows that the adoption of a change in coordinates
\begin{eqnarray}
R & = & A e^\eta\sinh\chi \\ \nonumber
T & = & A e^\eta\cosh\chi
\label{conf}
\end{eqnarray}
where $\eta$ is the conformal time, defined such that $dt=R_0ad\eta$, 
and $\chi = x / R_0$, then the FLRW metric can be written as
\begin{equation}
ds^2 = \frac{ R_0^2\ a^2(\eta)}{T^2 - R^2} \left[ dT^2 - dR^2 - 
R^2\left( d\theta^2 + \sin^2\theta d\phi^2 \right)\right]  \nonumber
\end{equation}
which is just 
\begin{equation}
ds^2 = \frac{ R_0^2\ a^2(\eta)}{T^2 - R^2}\ ds^2_{flat}
\label{incoords}
\end{equation}
Hence, lightcones  plotted in $R$-$T$ coordinates will  be the classical
light  curves of special  relativity (see  Figure~\ref{fig1}. \citet{is}
demonstrate that the motion of  fundamental observers in the FLRW metric
($\chi$=constant)  are  still  mapped  onto straight  lines  in  $R$-$T$
coordinates  and   with  this  choice   of  coordinate  transformations,
\citet{2006astro.ph..1171C} demonstrates that  such lines posses a slope
of
\begin{equation}
\beta = \frac{dR}{dT} = \frac{R}{T} = \tanh\chi
\label{velocity}
\end{equation}
Hence,  the fundamental observers  have a  constant velocity  across the
$R$-$T$   plane   given  by   $\beta$,   where  $\beta\rightarrow1$   as
$\chi\rightarrow\infty$.   This  is  taken   to  be  evidence  that  the
coordinate velocity is always less than  the speed of light, so that the
relative motion  of the fundamental  observers is always  subluminal, no
matter their  separation.  In this manner, it  appears that superluminal
motion can be removed through a coordinate transformation.

\section{Interpretation}\label{interpretation}

How are  we to interpret  this conclusion?  Has  superluminal expansion,
and hence  the expansion of  space, been refuted?  The  argument against
superluminal  recession boils  down  to the  finding, through  conformal
transformations, that the coordinate velocity is subluminal in conformal
coordinates.   However, as  was shown  in Section  \ref{velocities}, all
FLRW  universes---even in  the original  coordinates---posses coordinate
velocities that are  subluminal. Of greater importance is  the mapping of
proper  velocity to  conformal  coordinates.  Since  spacetime has  been
sliced up  differently, the surfaces of  constant coordinate time---over
which  proper distance  is measured---have  been altered.   The critical
concern is  therefore how this  new proper distance changes  relative to
the    new   time    coordinate.     This   was    not   addressed    in
\citet{2006astro.ph..1171C}.

\begin{figure*}
\centerline{ \psfig{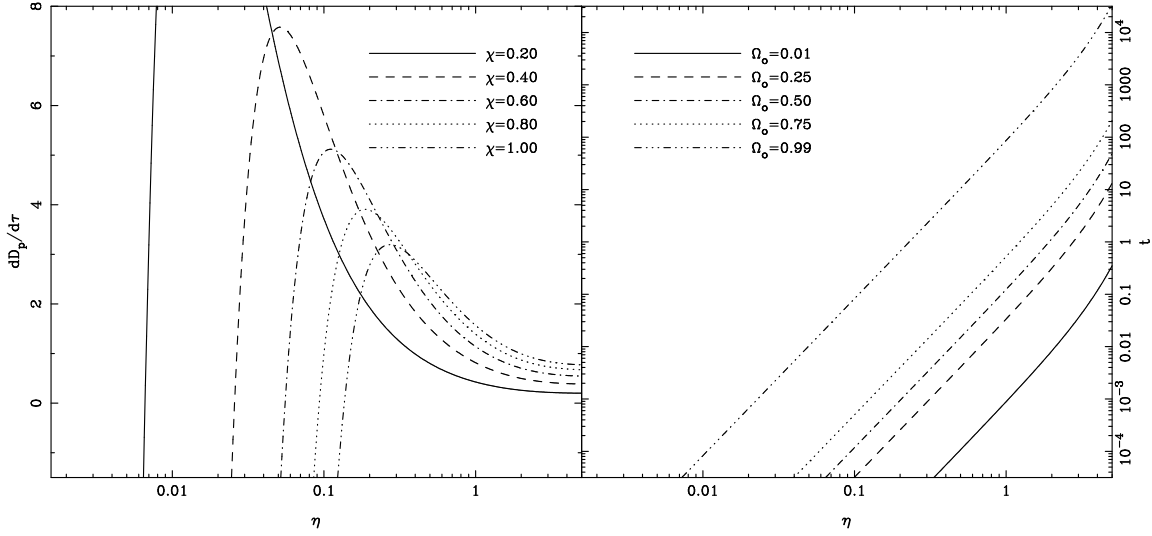}}
\caption[]{The left-hand panel presents the proper velocity in open
matter-only FLRW universes, for a range of conformal times $\eta$; clearly, at 
early times, the universe possess superluminal contraction and then
expansion. The right-hand panel presents the relationship between
the conformal time and the cosmic time of the standard FLRW universe
for a range of present day values of the matter density.
\label{fig3}}
\end{figure*}

To answer this, it is useful  to examine the picture of the example open
universe in $R$-$T$  coordinates (Figure~\ref{fig1}).  As FLRW universes
are conformally flat, light cones in this picture are at $45^\circ$.  As
seen in the coordinate  transformation given in Equation~\ref{conf}, all
fundamental   observers  (constant   $\chi$)  sit   on   straight  lines
originating at the  origin; note that the entire  (infinite) universe is
contained within the  outer lightcone. It might be  tempting to consider
the point  at $(R,T) =  (0,0)$ as  the FLRW Big  Bang, but in  fact this
`point' ($\eta=0$) is mapped to a hyperbola in the plane, from which the
paths of fundamental  observers extend, paths behind this  curve have no
physical equivalent in the FLRW universe.

What do we mean when we say the Universe is expanding?  It does not mean
that coordinates  are changing  in some particular  fashion, as  even in
standard FLRW  universe, objects maintain  spatial coordinate separation
(i.e.\ the fundamental or comoving spatial coordinates are separate). In
fact, universal  expansion should be  interpreted as an increase  in the
physical separation of objects with cosmic time i.e.\ a galaxy at $B$ is
moving  away from  $A$ at  so many  metres per  second, with  time being
measured by $A$'s clock, and distance being the proper distance.

\citet{2006astro.ph..1171C}  notes  that   for  a  spatially  flat  FLRW
universe, the conformal representation is
\begin{equation}
ds^2 = a^2(\eta)(d\eta^2 - d\chi^2 - \chi^2 \left( d\theta^2 + \sin^2\theta
  d\phi^2 \right) ),
\end{equation}
so that the distance to a  galaxy at comoving coordinate $\chi=X$ from a
fundamental  observer  at $\chi=0$  is  taken  along  a hypersurface  of
constant cosmological time ($d\eta=0$) and is
\begin{equation}
D_p(\eta) = \int \sqrt{ -ds^2 } = a(\eta)\int_0^X d\chi = a(\eta) X
\end{equation}
whereas the  proper time, $\tau$ as  measured by the  fundamental observer at
the origin {\rref is related to the coordinate time $t$ and conformal time
$\eta$ via}
\begin{equation}
d\tau = dt = a(\eta) d\eta.
\end{equation}
{\rref The rate of change of the proper distance to a comoving observer at 
$\chi=X$ in terms of the proper time
as measured at the origin is}
\begin{equation}
\frac{dD_p}{d\tau} = \frac{1}{a}\frac{da}{dt} (a X)
\end{equation}
{\rref For a flat universe, the radial coordinate $X$ is unbound and} 
hence, even in this conformal representation, superluminal expansion remains a
feature.

What about the conformal  representation of the open universe considered
by \citet{2006astro.ph..1171C}?  As  this is a coordinate transformation
from the FLRW universes, the distance is a line integral
\begin{equation}
D_p(\eta) = \int \sqrt{ -ds^2 } = R_0 a(\eta)\int \frac{\sqrt{dR^2
-dT^2}}{\sqrt{T^2-R^2}},
\end{equation}
with the  condition that  the path  be restricted to  a hyperbola  in the
$R$-$T$  plane ($\eta$=constant),  so  that $T^2  -  R^2 =  A^2e^{2\eta}
\equiv  k^2$.  From  this  obtains  the relation  $dT  =  (R/T)dR$;  the
integration proceeds from the origin along to a point $R(\chi)=R_\chi$:
\begin{equation}
  \frac{D_p(\eta)}{a(\eta)R_0} = \int_0^{R_\chi} \frac{dR}{T} = \int^{R_\chi}_0 \frac{dR}{\sqrt{ k^2 + R^2 }} =
  \mathrm{asinh}\left(\frac{R_\chi}{k}\right) = \chi
\end{equation}
This  physical separation---even  in this  conformal representation---is
that expected from the standard FLRW analysis.

But of course, one of the joys of relativity is the ability to slice and
dice spacetime  differently for differing observers, and  we can instead
calculate  the  distance  along  the spatial  hypersurfaces  defined  by
constant  $T$ in  the  conformal representation;  this  is the  approach
adopted by  \citet{2006astro.ph..1171C}.  Does this  remove superluminal
expansion?  Remembering that in an open, matter-only universe,
\begin{equation}
t(\eta) = \frac{\Omega_0}{2(1-\Omega_0)^{\frac{3}{2}}}
\left(\sinh\ \eta - \eta\right),
\nonumber
\end{equation}
\begin{equation} 
a(\eta) = \frac{\Omega_0}{2(1-\Omega_0)}\left(\cosh\ \eta - 1\right),
\label{cosmo}
\end{equation}
where   $\Omega_0$   is   the    present   day   normalized   matter   density
\citep[see][]{2005gere.book.....H}.     Hence,   the   distance    along   the
hypersurface is (taking $A=1$ for convenience)
\begin{eqnarray}
  D_p(T) & = & \frac{R_0 \Omega_0}{2(1-\Omega_0)} \int_0^R \frac{
    \cosh( ln( \sqrt{T^2-R'^2} )) - 1}{\sqrt{T^2-R'^2}} dR' \\ \nonumber
  & = & \frac{R_0 \Omega_0}{4(1-\Omega_0)} \left[ R - 2 \mathrm{atan}\left( \sinh\chi\right) + \frac{\chi}{T}\right].
\end{eqnarray}
Figure~\ref{fig2} presents  this proper distance  as a function of  the proper
time experienced  by an  observer at $R=0$  for three fiducial  universes with
$\Omega_0 = 0.01, 0.5$ and $0.99$.  In each, the solid {\rref lines represent}
the  proper distance,  while the  dashed  {\rref lines}  at $45^\circ$  {\rref
  represent}  the speed  of  light.  The dotted  {\rref  lines} represent  the
distance in terms of the  conformal coordinates while neglecting the conformal
factor outside the metric (i.e.\ over Minkowski spacetime).

For the  low density case, the  conformal factor tends to  unity and the
spacetime becomes that of special relativity. Hence, the proper distance
increases as  expected in this representation; the  paths are subluminal
and match  those calculated in  the $R$-$T$ coordinates. However,  as we
increase the mass density of the  universe, it is seen that the increase
of  the  proper  distance  with  proper  time  deviates  from  Minkowski
spacetime, in places  being superluminal.  This is very  apparent in the
case  where $\Omega_0=0.99$  where the  majority of  paths  are receding
superluminally.

It  is interesting  to  examine the  properties  of this  proper velocity  for
constant $T$  slices in a little  more detail.  {\rref Noting  that the proper
  time  $\tau$  for  an  observer  the  origin is  related  to  the  conformal
  coordinate time $T$ via}
\begin{equation}
d\tau = \frac{R_0 a(\eta)}{T} dT
\end{equation}
it  is straight forward to
show that
\begin{equation}
  \frac{dD_p}{d\tau} = 
  \frac{dT}{d\tau}\frac{dD_P}{dT} = 
\left[ e^\eta\ \tanh\ \chi -
    \frac{\chi}{e^\eta}\right]\frac{1}{\left( e^\eta + e^{-\eta}
      -2\right)}
\end{equation}
{\rref where $\eta$ is the conformal time ticked off at the origin and is 
  related to the proper time at the origin via $d\tau=R_0 a d\eta$}.
Importantly, the form  of the curve is independent of  $\Omega_0$ and hence is
valid for  all open ($0<\Omega_0<1$)  FLRW universes.  The left-hand  panel of
Figure~\ref{fig3}  presents this  function for  several values  of  $\chi$; as
$\eta\rightarrow\infty$, $dD_p/d\tau\rightarrow\  \tanh\ \chi$, the coordinate
velocity, but it is clear from this figure that at early times, the coordinate
velocity  is negative  and superluminal,  becoming subluminal  before becoming
positive and superluminal again; this is true for all values of $\chi$.

The remaining issue is the relation between the FLRW conformal time $\eta$ and
the  cosmological time  $t$;  this  is given  by  Equation~\ref{cosmo} and  is
presented  in the  right-hand panel  of Figure~\ref{fig3}.   As  expected from
Figure~\ref{fig2},  in  the   $\Omega_0=0.01$  universe,  the  conformal  time
approaches  5 in  a fraction  of a  Hubble time  (i.e.  $t<1$)  and  hence the
superluminal motion  occurred in the very  early universe and  is not apparent
given  the   resolution  of   Figure~\ref{fig2}.   However,  for   the  {\rref
  $\Omega_0=0.99$}  universe,  this  conformal  time  of  $\eta\sim5$  is  not
approached  until after  several  hundred Hubble  times  and the  superluminal
expansion is apparent over cosmic  history.  However, in the distant universe,
this superluminal  motion will  be lost  as the proper  velocity tends  to the
coordinate  velocity.  Note,  that as  $\Omega_0\rightarrow 0$,  the excessive
superluminal motion is pushed back to  earlier epochs of cosmic time $t$ until
$\Omega_0=0$, the  expansion is that of empty,  special relativistic universe,
with the same proper and coordinate velocity.

\section{Conclusions}\label{conclusions}
In short, a recent interpretation of  the nature of the expansion of the
universe in conformal coordinates concludes that superluminal expansion,
a  staple of  FLRW  universes, is  nothing  but a  coordinate effect  of
general relativity  and it  can be removed  through a  simple coordinate
transformation.  This paper  has examined this claim and  has found this
conclusion  to  be erroneous  and  objects  in  the universe  can  still
physically  separate  at  superluminal  velocities,  even  in  conformal
coordinates.  It  should be noted  that the incorrect  interpretation of
conformal coordinates is not new; \citet{1998ApJ...508..129Q} attacked a
series of  papers which claimed  cosmology in conformal  coordinates can
even      remove       the      need      for       a      big      bang
\citep{1994ApJ...434..397E,1995MNRAS.277..627E,1997ApJ...479...40E}.  As
ever, in relativity,  one should be careful about  the interpretation of
coordinates and the definition of distances.

In  a companion  paper,  \citet{Francis} discussed  a  number of  issues
relating to the  recent discussions on the meaning  and use of expanding
space as a concept in cosmology,  and we reiterate the most important of
these now.   The FLRW metric  of the cosmos  contains a term,  the scale
factor, which grows with time  in an expanding universe. It is perfectly
acceptable to talk  of this metric expansion as  the expansion of space,
but ones intuition must be lead by the mathematical framework of general
relativity. If, however,  one wishes to adopt the  conformal metric with
the flat spacetime of  special relativity (although a changing conformal
factor  in front  of it),  that is  equally acceptable.   The  choice of
coordinates is down  to personal preference, as both  must give the same
predictions.  From all of this, it  should be clear that it is futile to
ask the question ``is space {\it really} expanding?''; the standard-FLRW
metric  and its  conformal representation  are the  same  spacetime.  No
experiment  can be formulated  to differentiate  one personal  choice of
coordinates from another.

\section*{Acknowledgments}
The anonymous referee  is thanked for their comments  on this manuscript.  GFL
acknowledges support from ARC Discover Project DP0665574.  GFL also thanks the
possums that  fight on  his back  deck at 4am,  waking him  up and  giving him
plenty of time to  think before his kids wake up at  6am, and also thanks Bryn
and Dylan for reintroducing him to {\it Jason and the Argonauts}.

\end{document}